\documentclass[]{article}
\usepackage[a4paper, left=2cm, right=2cm, top=2cm, bottom=2cm]{geometry}
\usepackage[version=3]{mhchem}
\usepackage[utf8]{inputenc}
\usepackage{booktabs}
\usepackage{graphicx}
\usepackage{caption}
\usepackage{subcaption}
\usepackage{changepage}
\usepackage{gensymb}
\usepackage{cite}
\usepackage{datetime}
\usepackage{float}
\usepackage[british]{isodate}
\usepackage{enumerate}
\usepackage[space]{grffile}
\usepackage{setspace}
\usepackage{tabularx}
\usepackage{array,booktabs}
\pdfoutput=1

\title{Exact quantized momentum eigenvalues and eigenstates of a general molecular potential model}
\date{}
\author{Mahmoud Farout$^{1,*}$, Ahmed Bassalat$^{1,*}$ and Sameer M. Ikhdair$^{1,2}$}

\begin{document}

\maketitle

\noindent $^1$ Department of Physics, An-Najah National University, Nablus, Palestine \\
$^2$ Department of Electrical Engineering, Near East University, Nicosia, Northern Cyprus, Mersin 10, Turkey\\
$^*$ Corresponding authors:  ahmed.bassalat@najah.edu and
m.qaroot@najah.edu\\

\section*{\bf {\centering Abstract}}
\noindent We obtain the quantized momentum eigenvalues, $P_n$, and the momentum eigenstates for the space-like Schrodinger equation, the Feinberg-Horodecki equation, with the general molecular potential which is constructed by the temporal counterpart of the spatial form of these potentials. The present work is illustrated with  two special cases of the general form: time-dependent Wie-Hua Oscillator and time-dependent Manning-Rosen potential. We also plot the variations of the general molecular potential with its two special cases and their momentum states for few quantized states against the screening parameter. 
\\

\textbf{Keywords:} Bound states; Feinberg-Horodecki equation; the time-dependent general molecular potential; time-dependent Wie-Hua Oscillator; and time-dependent Manning-Rosen potential.\\

PACS:03.65.-W;03.65.Pm.

\section{Introduction}
Any physical phenomenon in nature is usually characterized by solving differential equations. The time-dependent
Schrödinger equation represents an example that describes quantum-mechanical phenomena, in which it dictates the dynamics of a quantum system. Solving this differential equation by means of any method results in the eigenvalues and eigenfunctions of that Schrödinger quantum system. However, solving time-dependent
Schrödinger equation analytically is not easy except when the time-dependent potentials are constant, linear and quadratic functions of the coordinates \cite{1,2,3,4}.
The Feinberg-Horodecki (FH) equation is a space-like counterpart of the Schrödinger equation which was derived by Horodecki \cite{5} from the relativistic Feinberg equation \cite{6}. This equation has been demonstrated in the possibility of describing biological systems \cite{7, 8} in terms of the time-like supersymmetric quantum mechanics \cite{9}. The space-like solutions of the FH equation can be employed to test its relevance
in different areas of science including physics, biology and medicine \cite{7, 8}. Molski constructed the space-like coherent states of a time-dependent Morse oscillator on the basis of the FH quantal equation to minimize the uncertainty in the time-energy relation and showed that the results are useful for interpreting the formation of the
specific growth patterns during the crystallization process and the growth in biological systems \cite{7}. In addition, Molski obtained FH equation to demonstrate a possibility of describing the biological systems in terms of the space-like quantum
supersymmetry for an-harmonic oscillators \cite{8}.

Recently, Bera and Sil found the exact solutions of the FH equation for the time-dependentWei-Hua oscillator
and Manning-Rosen potentials by the Nikiforov-Uvarov (NU) method \cite{10}. In 1957, Deng and Fan \cite{11}
proposed a potential model for diatomic molecules named as the Deng-Fan oscillator potential. This potential is also known as general Morse potential \cite{12,13} whose analytical expressions for energy levels and wave functions have been derived \cite{11, 12, 13, 14} and related to the Manning-Rosen potential \cite{15, 16} (also called
Eckart potential by some authors \cite{17, 18, 19} is anharmonic potential. It obeys the correct physical boundary conditions at t = 0 and 1. The space-like Deng-Fan potential is qualitatively similar to the Morse potential but has the correct asymptotic behavior when the inter-nuclear distance goes to zero \cite{11} and used to describe ro-vibrational energy levels for the diatomic molecules and electromagnetic transitions \cite{20, 21, 22}.
The exact momentum state solutions of the FH equation with the rotating time-dependent Deng-Fan oscillator
potential are presented within the framework of the generalized parametric NU method. The energy
eigenvalues and corresponding wavefunctions are obtained in a closed form \cite{23}.

Recently, Altug and Sever have studied the FH equation with time-dependent Poschl-Teller potential and
found its space-like coherent states \cite{24}. We also studied the solutions of FH equation for time-dependent mass (TDM) harmonic oscillator quantum system. A certain interaction is applied to a time-dependent mass m(t) to provide a particular spectrum of stationary energy. The  spectrum related to the Harmonic oscillator potential acting on the TDM stationary state energies is found \cite{25}. The exact solutions of FH equation under time-dependent Tietz-Wei di-atomic molecular potential have been obtained. In particular, the quantized momentum eigenvalues and
corresponding wave functions are found in the framework of supersymmetric quantum mechanics \cite{26}. The spectra of general molecular potential (GMP) are obtained using the asymptotic iteration method within the framework of
non-relativistic quantum mechanics. The vibrational partition function is calculated in closed form and used to
obtain thermodynamic functions \cite{27}.

In new work, we have obtained the quantized momentum solution of the FH equation with combined Kratzer plus screened Coulomb potential using NU method. We constructed three special cases of this general form; the time-dependent modified Kratzer potential, the time-dependent screened Coulomb potential and the time-dependent Coulomb potential \cite{28}.

The motivation of this work is to apply the Nikiforov-Uvarov method \cite{29} for the general molecular potential having a certain time-dependence. The momentum eigenvalues, $P_n$, of the FH equation and the space-like coherent eigenvectors are obtained. The rest of this work is organized as follows: the NU method is briefly introduced in Section 2. The exact solution of the FH equation for the time-dependent general molecular potential is solved to obtain its quantized momentum states and eigenfunctions in Section 3. 
We generate the solutions of a few special potentials mainly found from our general form solution in section 4. Finally we present our discussions and conclusions.

\section{Exact solutions of the FH equation for the time-dependent general molecular potential }
The Nikiforov-Uvarov (NU) method (see appendix \ref{Appendix}) will be applied to find the exact solutions of FH equation for the general molecular potential then the eigenvalues and eigenfunctions of two special cases are produced from the results.

The time-dependent of the general molecular potential is given by \cite{30}
\begin{equation}
V(t)= \frac{A- B e^{-\alpha(t-t_e)}+ \tilde{q}[C-D e^{-\alpha(t-t_e)}]^2}{[1-qe^{-\alpha(t-t_e)}]^2},
\label{eq: GMP}
\end{equation}
where $A$, $B$, $C$, $D$, and $\alpha$  are adjustable real potential parameters. $\tilde{q}$ and $q$ are dimensionless parameters. $t_e$ the equilibrium time point. The parameters of the potential must satisfy the condition $\frac{1}{\alpha} ln q+ t_e \leq 0$ to avoid singularity.
If the general potential is substituted in FH equation, one obtains
\begin{equation}
\left[-\frac{\hbar^2}{2mc^2}\frac{d^2}{dt^2}+ \left(\frac{A- B e^{-\alpha(t-t_e)}+ \tilde{q}[C-D e^{-\alpha(t-t_e)}]^2}{[1-qe^{-\alpha(t-t_e)}]^2}\right)\right]\psi_n(t) = cP_n \psi_n(t).
\end{equation}
Now, let $s=qe^{-\alpha(t-t_e)}$, where s $\epsilon$($0$, $qe^{\alpha t_e}$), we get
\begin{equation}
\frac{d^2\psi_n(s)}{ds^2} + \frac{(1-s)}{s(1-s)}\frac{d\psi_n(s)}{ds}+\frac{-\gamma_1^2-\gamma_3s+\gamma_2s^2}{s^2(1-s)^2}\psi_n(s)=0,
\label{eq:differential eq to solve}
\end{equation}
where 
\begin{equation}
\gamma_1^2=\frac{2mc^2}{\hbar^2 \alpha^2}\left(A+ \tilde{q}C^2-cP_n\right),
\label{eq:gamma1}
\end{equation}
\begin{equation}
\gamma_2=-\frac{2mc^2}{\hbar^2 \alpha^2}\left(\tilde{q}\frac{D^2}{q^2}-cP_n\right),
\end{equation}
\begin{equation}
\gamma_3=-\frac{2mc^2}{\hbar^2 \alpha^2}\left(\frac{B}{q}+2\tilde{q} C\frac{D}{q}-2cP_n\right).
\end{equation}
After comparing equation (\ref{eq:differential eq to solve}) with equation (\ref{eq: NU-equ}), one obtains \\
$\tilde{\tau}(s)=1-s$, $\sigma(s)= s(1-s)$, and $\tilde{\sigma}(s)=-\gamma_1^2-\gamma_3s+\gamma_2 s^2$.\\
When these values are substituted in equation $\pi(s)= \frac{\sigma^{'}-\tilde{\tau}}{2}\pm \sqrt{(\frac{\sigma^{'}-\tilde{\tau}}{2})^2-\tilde{\sigma}+ k\sigma}$ \cite{29}, we get
\begin{equation}
\pi(s)= -\frac{s}{2}\pm \sqrt{\left(\frac{1}{4}-\gamma_2-k\right)s^2+ (k+\gamma_3)s+\gamma_1^2}.
\label{eq: pi solving}
\end{equation}
As mentioned in the NU method, the discriminant under the square root, in equation (\ref{eq: pi solving}), has to be zero, so that the expression of $\pi (s)$ becomes the square root of a polynomial of the first degree. This condition can be written as
\begin{equation}
\left(\frac{1}{4}-\gamma_2-k\right)s^2+ (k+\gamma_3)s+\gamma_1^2=0.
\end{equation}
After solving this equation, we get
\begin{equation}
s= \frac{-(k+\gamma_3)\pm\sqrt{(k+\gamma_3)^2-4\gamma_1^2(\frac{1}{4}-\gamma_2-k)}}{2(\frac{1}{4}-\gamma_2-k)}.
\label{eq: s}
\end{equation}
Then, for our purpose we assume that
\begin{equation}
(k+\gamma_3)^2-4\gamma_1^2\left(\frac{1}{4}-\gamma_2-k\right)=0.
\end{equation}
Arranging this equation and solving it to get an expression for k which is given by the following,
\begin{equation}
k_\pm = -\gamma_3-2\gamma_1^2\pm 2\gamma_1\left(\frac{1}{R}-\frac{1}{2}\right),
\end{equation}
where the expression between the parentheses is given by
\begin{equation}
\frac{1}{R}-\frac{1}{2}= \sqrt{\frac{2mc^2}{\hbar^2 \alpha^2}\left[\tilde{q}\left(C-\frac{D}{q}\right)^2+A-\frac{B}{q}\right]+\frac{1}{4}}.
\label{eq:1/R-1/2}
\end{equation}
where the parameters in this equation must be selected to let R be real and the results have physical meanings. 
If we substitute $k_-$ into equation (\ref{eq: pi solving}) we get a possible expression for $\pi(s)$, which is given by 
\begin{equation}
\pi(s) =\gamma_1 (1-s) -\frac{s}{R},
\label{eq: pi(s) solution}
\end{equation}
this solution satisfy the condition that the derivative of $\tau(s)$ is negative. Therefore, the expression of $\tau(s)$ which satisfies these conditions can be written as
\begin{equation}
\tau(s) =1-s +2\gamma_1-2s(\gamma_1+\frac{1}{R}).
\label{eq: tau(s) result}
\end{equation}
Now, substituting the values of $\tau^{'}_-(s)$, $\sigma^{''}(s)$, $\pi_-(s)$ and $k_-$ into equations (\ref{eq: lambda1}) and (\ref{eq: lambda2}), we obtain
\begin{equation}
\lambda_n= \frac{2mc^2}{\hbar^2 \alpha^2}\left(\frac{B}{q}+2 \tilde{q}C\frac{D}{q}-2A-2\tilde{q}C^2\right)-\frac{2\gamma_1}{R}-\frac{1}{R},
\label{eq:lambda_n solution}
\end{equation}
and 
\begin{equation}
\lambda= \lambda_n= n(n+ \frac{2}{R})+ 2n \gamma_1.
\label{eq:lambda-solution}
\end{equation}
Now, from equations (\ref{eq:lambda_n solution}) and (\ref{eq:lambda-solution}), we get the eigenvalues of the quantized momentum as
\begin{equation}
P_n= \frac{1}{c}\left(A+\tilde{q}C^2-\frac{\alpha^2 \hbar^2}{2mc^2}\left[\frac{n(n+\frac{2}{R})+\frac{1}{R}-\frac{2mc^2}{\alpha^2 \hbar^2}(\frac{B}{q}+2\tilde{q}C\frac{D}{q}-2A-2\tilde{q}C^2)}{2(n+\frac{1}{R})}\right]^2\right).
\label{eq:eigenvalues of Pn}
\end{equation}
Due to the NU method used in getting the eigenvalues, the polynomial solutions of the hypergeometric function $y_n(s)$ depend on the weight function $\rho(s)$ which can be determined using NU procedure to get 
\begin{equation}
\rho(s)= s^{2\gamma_1}(1-s)^{(\frac{2}{R})-1}.
\label{eq:rho-result}
\end{equation}
Substituting the result of $\rho(s)$ into equation $y_n$ in \cite{28}, we get an expression for the wave functions as
\begin{equation}
y_n(s)= A_n s^{-2\gamma_1}(1-s)^{-(\frac{2}{R}-1)}\frac{d^n}{ds^n}\left[s^{n+2\gamma_1}(1-s)^{n+\frac{2}{R}-1}\right],
\label{eq: yn 2}
\end{equation}
where $A_n$ is the normalization constant. Solving equation (\ref{eq: yn 2}) gives the final form of the wave function in terms of the Jacobi polynomial $P_n^{(\alpha, \beta)}$ as follows,
\begin{equation}
y_n(s)= A_n n! P_n^{(2\gamma_1, \frac{2}{R}-1)}(1-2s).
\label{eq: yn last solution}
\end{equation}
Now, substituting $\pi_-(s)$ and $\sigma(s)$ into $\sigma(s)= \pi(s) \frac{\phi_n(s)}{{\phi_n}^{'}(s)}$ then solving it we obtain
\begin{equation}
\phi_n(s)= s^{\gamma_1}(1-s)^\frac{1}{R}.
\label{eq: phi solutin}
\end{equation}
Substituting equations (\ref{eq: yn last solution}) and (\ref{eq: phi solutin})in equation(\ref{eq: psi}), one obtains,
\begin{equation}
\psi_n(s)= B_n s^{\gamma_1}(1-s)^\frac{1}{R} P_n^{(2\gamma_1, (\frac{2}{R})-1)}(1-2s),
\end{equation}
where $B_n$ is the normalization constant. 
\section{Special cases}
\noindent Several potentials were proposed to obtain information about diatomic and polyatomic molecules structures. These potentials are represented by Rosen–Morse or Trigonometric P\"oschl–Teller potentials which are used for some vibrations of some polyatomic molecules such as NH$_3$ and SO$_2$ \cite{27}. The Tietz-Wei diatomic molecular potential was proposed as an inter-molecular potential and is considered as one of the best potential models which describes the vibrational energy of a diatomic molecules \cite{31, 32, 33, 34}. In addition, Manning-Rosen and Wie-Hua potentials have been proposed for diatomic molecule structure \cite{16, 33} and are discussed here as special cases of the general molecular potential.
\subsection{Time-dependent Wie-Hua oscillator}
\noindent A four-parameter potential function was introduced for bond-stretching vibration of diatomic molecules. It may fit the experimental RKR (Rydberg-Klein-Rees) curve more closely than the Morse function, especially when the potential domain extends to near the dissociation limit. The corresponding Schrödinger equation was solved exactly for zero total angular momentum and approximately for nonzero total angular momentum \cite{33}.
To get the Wie-Hua potential from the general form of the diatomic molecules potential, A= B=0, and C=D=1 are substituted in (\ref{eq: GMP}) to reduce the general form to the special case \cite{33},
\begin{equation}
V(t)=\tilde{q}\left(\frac{1-e^{-\alpha(t-t_e)}}{1-qe^{-\alpha(t-t_e)}}\right)^2.
\end{equation}
And by substituting the same constants in (\ref{eq:eigenvalues of Pn}) we get the eigenvalues of the time-dependent HF equation with Wie-Hua potential. The result is as follows 
\begin{equation}
P_n=\frac{1}{c}\left(\tilde{q}-\frac{\alpha^2 \hbar^2}{2mc^2}\left[\frac{n(n+\frac{2}{R})+\frac{1}{R}-\frac{2mc^2}{\alpha^2 \hbar^2}(\frac{2\tilde{q}}{q}-2\tilde{q})}{2(n+\frac{1}{R})}\right]^2\right),
\end{equation}
with
\begin{equation}
\frac{1}{R}=\frac{1}{2}+\sqrt{\frac{2mc^2 \tilde{q}}{\hbar^2 \alpha^2 }(1-\frac{1}{q})^2+\frac{1}{4}}.
\end{equation}
where $\alpha$, $\tilde{q}$ and $q$ must be selected to make R be real. 

To determine the eigenfunctions associated with the Wei-Hua potential, the same parameters were substituted in (\ref{eq:gamma1}) which results in 
\begin{equation}
\psi_n(s)= B_n (qe^{-\alpha(t-t_e)})^{\gamma_1}(1-(qe^{-\alpha(t-t_e)}))^\frac{1}{R} P_n^{(2\gamma_1, (\frac{2}{R})-1)}(1-2(qe^{-\alpha(t-t_e)})),
\end{equation}
where 
\begin{equation}
\gamma_1=\frac{n(n+\frac{2}{R})+\frac{1}{R}-\frac{4mc^2\tilde{q}}{\alpha^2\hbar^2}(\frac{1}{q}-1)}{2(n+\frac{1}{R})},
\end{equation}
which agrees with the result obtained in  \cite{10}.
\subsection{Time-dependent Manning–Rosen potential}
\noindent This potential is used as a mathematical model in describing diatomic molecular vibrations. It is also employed in several branches of physics in studying their bound states and scattering properties. 
It is well known that Schr\"{o}dinger equation can be solved exactly for this potential for s-wave i.e., $l=0$. But, for an arbitrary l-states, i.e.,  $l$ is not equal to $0$, the Schrödinger equation cannot be solved exactly. Therefore, the Schrödinger equation is solved numerically or approximately using approximation schemes. Some authors used the approximation scheme proposed by Greene and Aldrich \cite{35} to study analytically the $l \neq 0$ bound states or scattering states of the Schrödinger or even relativistic wave equations for Manning-Rosen potential. We calculate and find its  quantized momentum states and normalized wave functions \cite{16}. 

By substituting $A=C=0$, $q=1$, $B=V_0 \alpha^2$ and $\tilde{q} D^2= \alpha^2(\beta(\beta-1)+V_0)$ in equation (\ref{eq: GMP}) we get the Manning-Rosen potential \cite{16},
\begin{equation}
V(t)=\frac{-V_0 \alpha^2 e^{-\alpha(t-t_e)} + \alpha^2(\beta(\beta-1)+V_0) e^{-2\alpha(t-t_e)}}{(1-e^{-\alpha(t-t_e)})^2},
\end{equation}
 and by substituting the values of $A$, $C$, $B$, $q$ and $\tilde{q} D^2$ in equation (\ref{eq:eigenvalues of Pn}) gives the eigenvalues of the FH time dependent equation. The eigenvalues are given by the relation,
\begin{equation}
P_n= - \frac{\hbar^2 \alpha^2}{2mc^3}\left[\frac{n\left(n+\frac{2}{R}\right)+\frac{1}{R}-\frac{2mc^2 (v_o \alpha^2)}{\hbar^2 \alpha^2 }}{2\left(n+\frac{1}{R}\right)}\right]^2,
\end{equation}
where 
\begin{equation}
\frac{1}{R}= \frac{1}{2} +\sqrt{\frac{2mc^2}{\hbar^2 }[\beta(\beta-1)]+\frac{1}{4}}.
\end{equation}

To determine the eigenfunctions associated with the Manning-Rosen potential, the same parameters were substituted in (\ref{eq:gamma1}) which results in 
\begin{equation}
\psi_n(s)= B_n (qe^{-\alpha(t-t_e)})^{\gamma_1}(1-(qe^{-\alpha(t-t_e)}))^\frac{1}{R} P_n^{(2\gamma_1, (\frac{2}{R})-1)}(1-2(qe^{-\alpha(t-t_e)})),
\end{equation}
where 
\begin{equation}
\gamma_1= \frac{n\left(n+\frac{2}{R}\right)+\frac{1}{R}-\frac{2mc^2 (v_o \alpha^2)}{\hbar^2 \alpha^2 }}{2\left(n+\frac{1}{R}\right)}.
\end{equation}
These results agree completely with that obtained in \cite{10}.
\section{Numerical Results and Discussion}
\noindent To illustrate the physical meaning of our investigation, we plot some figures of the potentials and the corresponding momentum obtained in each case versus time or the screening parameter by choosing suitable potential parameters. Figure 1 shows a plot of the general molecular potential against the oscillating time for various values of the screening parameter $\alpha$. For higher values of time, the potential tends to be constant when $\alpha$ takes larger values. 
In Fig. 2, we examine the variations in the FH quantized momentum, for various states, against the screening parameter $\alpha$. It is noted that as $\alpha$ increases, the momentum of the system increases from the negative region to the positive region. Obviously, the momentums for higher states are close to each other in positive region near $\alpha$=0.5. 
In Fig.3, it is seen that the quantized momentum of the system decreases monotonically as the potential strength parameter B becomes negative. Therefore, when  state n increases, the momentum decreases from positive region to negative region for higher values of $\alpha$. This behavior is opposite to Fig. 2, when potential strength parameter B is taken to positive. 

\begin{figure}[H]
	\includegraphics[width=1.05\linewidth]{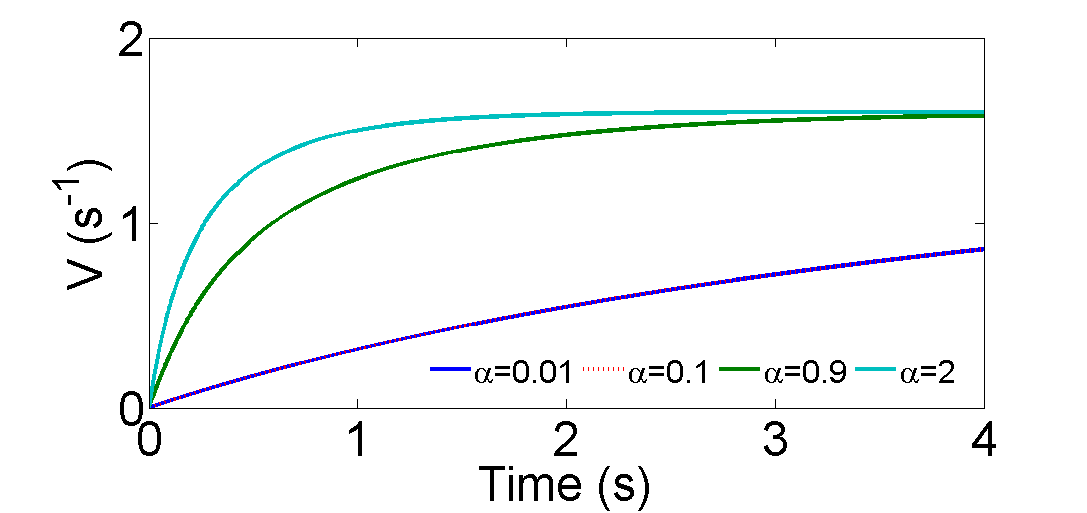}
	\caption[General Potential]{The general potential for diatomic molecules. $\mu=\hbar=c$=1, $A=B=q=0.6$, and $C=D=$ $\tilde{q}$=1.}
	\label{fig:general pot}
\end{figure}

\begin{figure}[H]
	\includegraphics[width=1.05\linewidth]{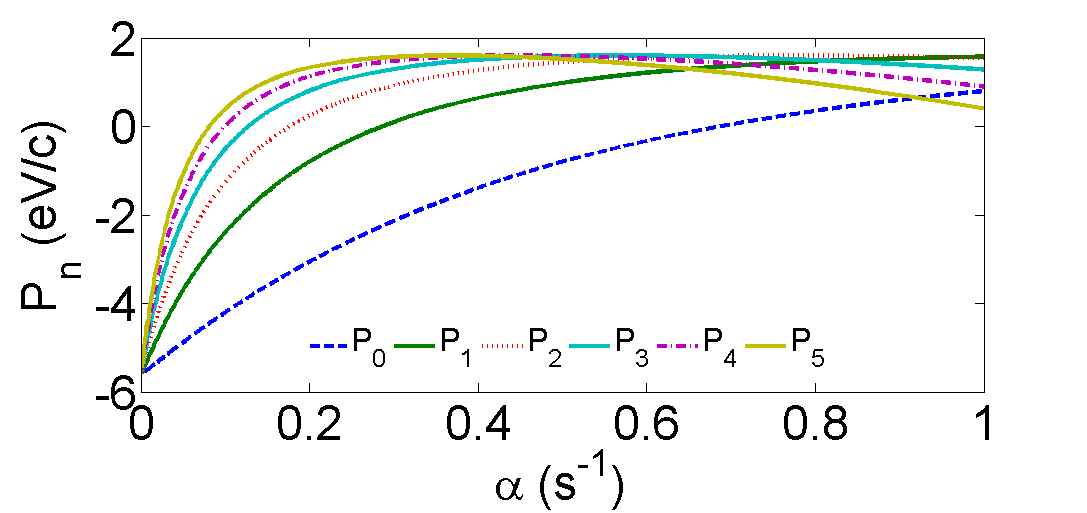}
	\caption[General Potential]{The FH quantized momentum eigenvalues of the general potential for diatomic molecules. $\mu=\hbar=c$=1, $A=B=q=0.6$, and $C=D=$ $\tilde{q}$=1.}
	\label{fig:general momentum_pos }
\end{figure}

\begin{figure}[H]
	\includegraphics[width=1.05\linewidth]{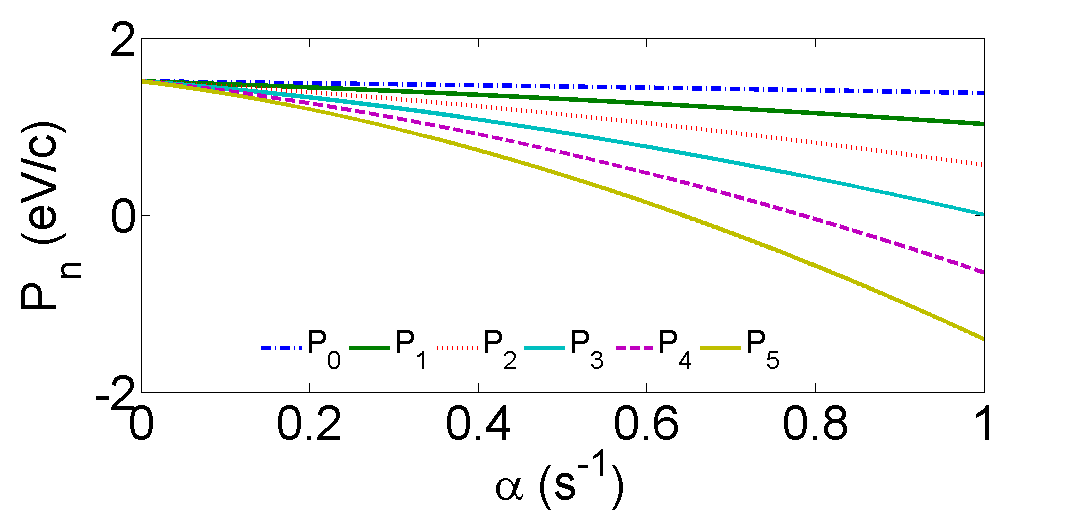}
	\caption[General Potential]{The FH quantized momentum eigenvalues of the general potential for diatomic molecules. $\mu=\hbar=c$=1, $A=q=0.6, B=-0.6$ and $C=D=$ $\tilde{q}$=1.}
	\label{fig:general momentum_neg }
\end{figure}
\begin{figure}[H]
	\includegraphics[width=1.05\linewidth]{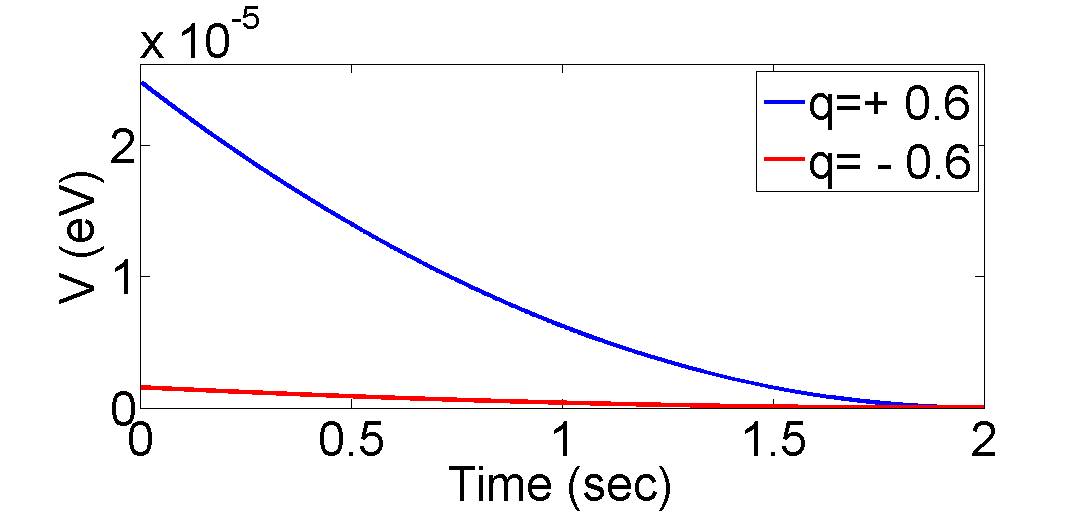}
	\caption[General Potential]{The Wie-Hua potential for diatomic molecules. $\mu=\hbar=c$=1, $A=B=0$, $\alpha=0.01$, and $C=D=$ $\tilde{q}$=1.}
	\label{fig:Wie pot}
\end{figure}
Figure 4 shows the variation of the time-dependent Wie-Hua potential for diatomic molecules, when potential strength parameters $A=B=0$, against time for two values of $q$. It is clear that decay is fast when $q>0$ whereas, decay is slow when $q<0$. 

Figure 5 plots the quantized momentum states versus the screening parameter $\alpha$ for $q<0$. This indicates that momentum decreases monotonically in the negative region when state increases and hence momentum states scatter away with increasing $\alpha$. A reverse behavior is shown in Fig. 6 when the quantized momentum states are plotted against $\alpha$ for $q>0$. The momentum state increases as n increases in positive region (up to $\alpha=0.3$). 
\begin{figure}[H]
	\includegraphics[width=1.05\linewidth]{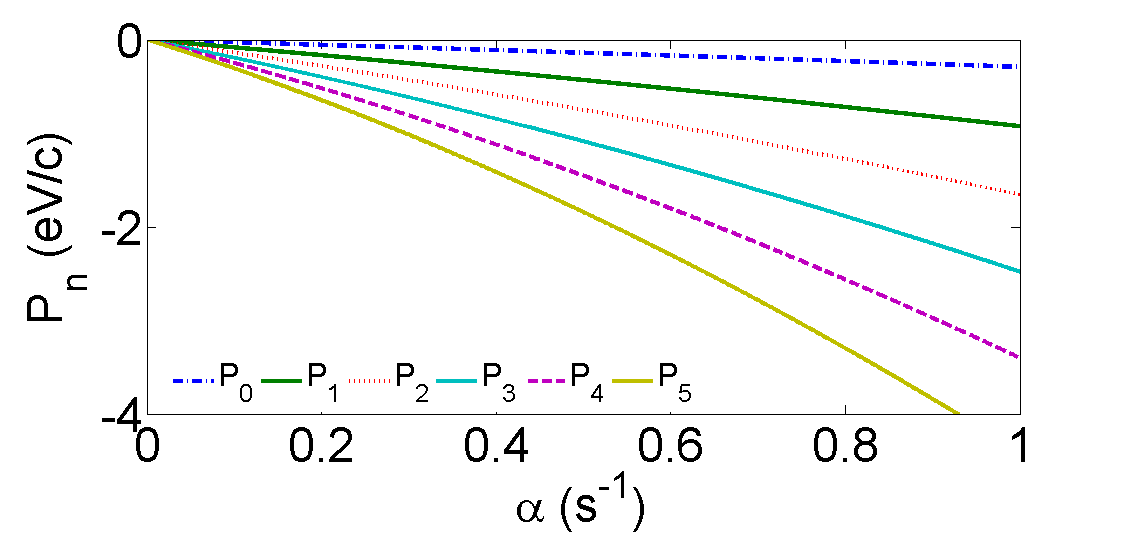}
	\caption[General Potential]{The FH quantized momentum eigenvalues of the Wie-Hua potential for diatomic molecules. $\mu=\hbar=c$=1, $q=- 0.6$, $A=B=0$ and $C=D=$ $\tilde{q}$=1.}
	\label{fig:Wie momentum_neg }
\end{figure}

\begin{figure}[H]
	\includegraphics[width=1.05\linewidth]{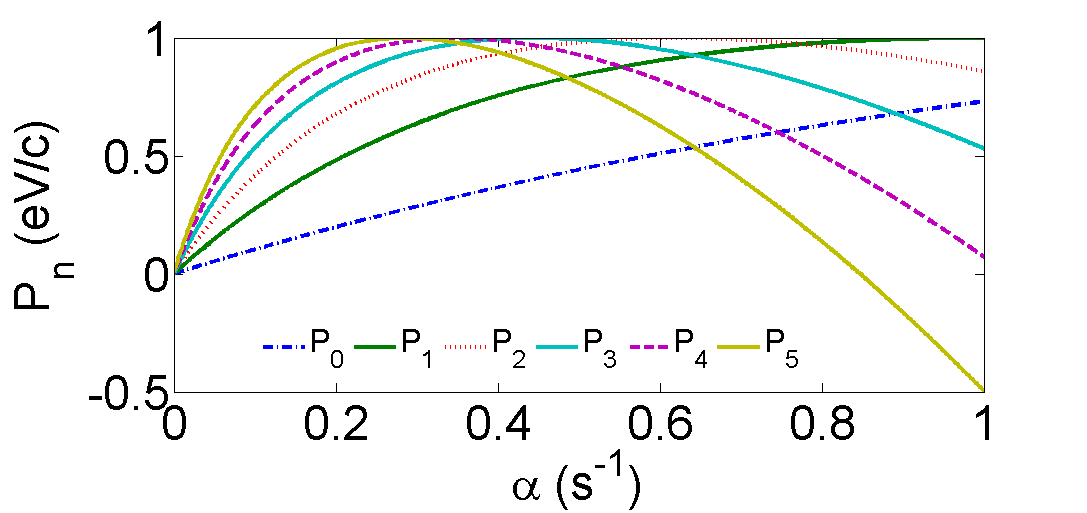}
	\caption[General Potential]{The FH quantized momentum eigenvalues of the Wie-Hua potential for diatomic molecules. $\mu=\hbar=c$=1, $q= 0.6$, $A=B=0$ and $C=D=$ $\tilde{q}$=1.}
	\label{fig:Wie momentum_pos }
\end{figure}
Figure 7 shows the time-dependent Manning-Rosen potential for diatomic molecules against oscillating time for two values of screening parameter $\alpha$. It increases with increasing values of $\alpha$ but the potential remains in the negative region (bound). 
In Fig. 8, we show the behavior of the quantized momentum states against $\alpha$ for the time-dependent intermolecular Manning-Rosen potential. When the potential strength parameter $V_0>0$, an increase in state n results in a monotonic decrease in momentum,  as $\alpha$ increases. Figure 9, a case in which $V_0<0$, shows the same behavior as in Fig. 8, but momentum is strongly more negative than before.
\begin{figure}[H]
	\includegraphics[width=1.05\linewidth]{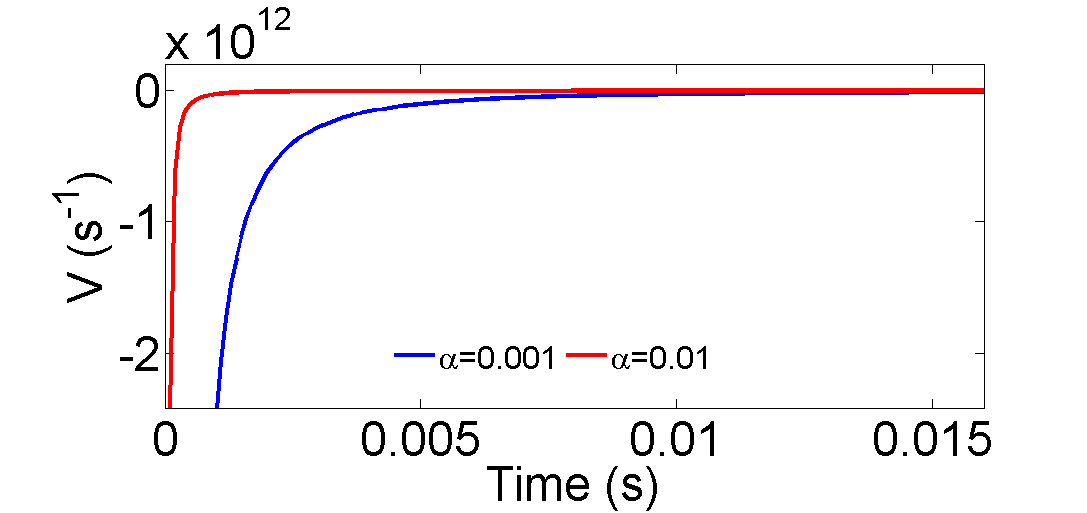}
	\caption[General Potential]{The Manning-Rosen potential for diatomic molecules. $\mu=\hbar=c$=1, $A=C=0$, $V_0$=2.5, $\beta=5$, and $q=$ $\tilde{q}$=1.}
	\label{fig:Manning pot}
\end{figure}

\begin{figure}[H]
	\includegraphics[width=1.05\linewidth]{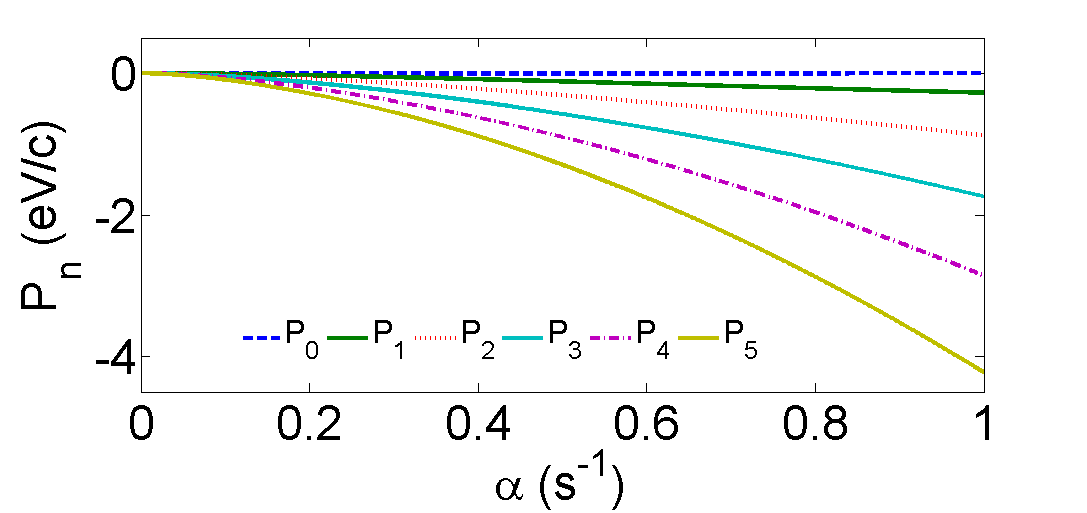}
	\caption[General Potential]{The FH quantized momentum eigenvalues of the Manning-Rosen potential for diatomic molecules. $\mu=\hbar=c$=1, $A=C=0$, $V_0$=2.5, $\beta=5$, and $q=$ $\tilde{q}$=1.}
	\label{fig:Manning momentum_pos}
\end{figure}

\begin{figure}[H]
	\includegraphics[width=1.05\linewidth]{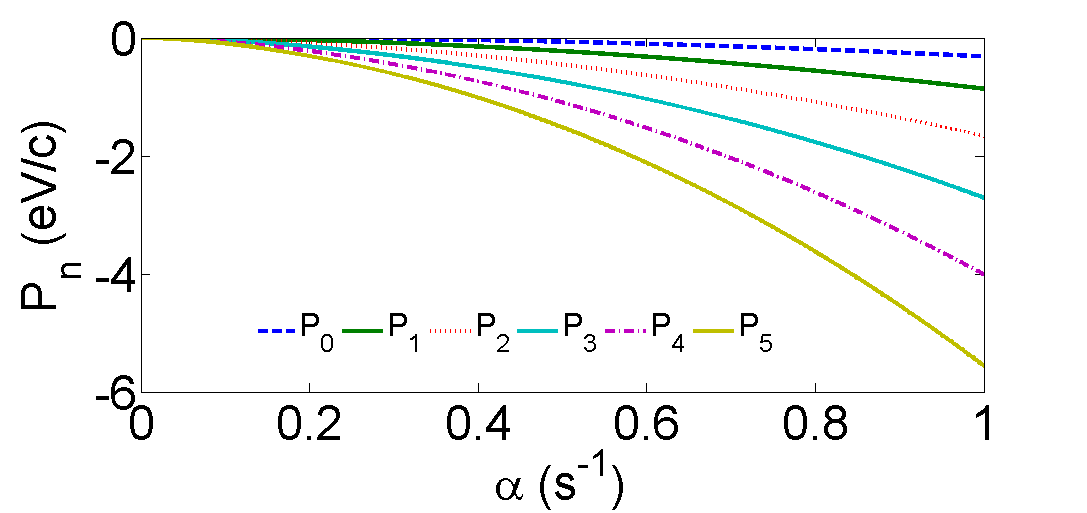}
	\caption[General Potential]{The FH quantized momentum eigenvalues of the Manning-Rosen potential for diatomic molecules. $\mu=\hbar=c$=1, $A=C=0$, $V_0$= -2.5, $\beta=5$, and $q=$ $\tilde{q}$=1.}
	\label{fig:Manning momentum_neg}
\end{figure}

\section{Conclusions}
\noindent We solved the Feinberg-Horodecki (FH) equation for the time-dependent general molecular potential via Nikiforov-Uvarov (NU) method. We got the exact quantized momentum eigenvalues solution of the FH equation. It is therefore, worth mentioning that the method is elegant and powerful. Our results can be applied in biophysics and other branches of physics.
In this paper, we have applied our result for the Wie-Hua and the Manning-Rosen potentials, as special cases of the general molecular potential, for quantized momentum eigenvalues. 
We find that our analytical results are in good agreement with other findings in literature,
The quantized momentum eigenvalues and their corresponding eigenfunctions are obtained exactly for the two exactly solvable problems. We have shown the behaviors of the general molecular potential as well as the two special cases. namely, Wie-Hua and manning-Rosen potentials against screening parameters.
Further, taking spectroscopic values for the potential parameters, we plotted the quantized momentum of few states against the screening parameter for diatomic molecules. Our results are good agreements with the energy bound states.

\appendix
\section{Appendix: Nikiforov-Uvarov method}
\label{Appendix}
In this section, we are briefly reviewing the Nikiforov-Uvarov (NU) method \cite{29}. Here we introduce the main points leaving the details into \cite{28}. The NU method is usually employed in reduction of the given second-order differential, which we are dealing with, into a general form of a hypergeometric type by using an appropriate coordinate transformation, $s=s(r)$, into the following standard form:
\begin{equation}
\psi_n^{''}(s)+ \frac{\tilde{\tau}(s)}{\sigma(s)} \psi_n^{'}(s)+ \frac{\tilde{\sigma}(s)}{\sigma^2(s)} \psi_n(s)=0,
\label{eq: NU-equ}
\end{equation}
where $\sigma (s)$ and $\tilde{\sigma}(s)$ are polynomials, of second-degree or less, and $\tilde{\tau}(s)$ is a first-degree polynomial. The wave function takes the form,
\begin{equation}
\psi_n(s)= \phi_n(s) y_n(s),
\label{eq: psi}
\end{equation}
which transforms equation (\ref{eq: NU-equ}) into a hypergeometric of the form
\begin{equation}
\sigma(s) y_n^{''}(s) + \tau(s) y_n^{'}(s) + \lambda y_n(s)= 0,
\label{eq:hypergeometric}
\end{equation}
where $\lambda$ in equation (\ref{eq:hypergeometric}) is a parameter defined as,
\begin{equation}
\lambda = \lambda_n = -n \tau^{'}(s) - \frac{n(n-1)}{2} \sigma^{''}(s),
\label{eq: lambda1}
\end{equation}
and $\lambda$ in equation (\ref{eq:hypergeometric}) is also defined as,
\begin{equation}
\lambda = \lambda_n= k+ \pi^{'}(s),
\label{eq: lambda2}
\end{equation}
These two definitions are exploited to calculate the eigenvalues of the system. More details are left to the reader in Ref. \cite{28}. 

\end{document}